\documentclass[acmlarge]{acmart}

\usepackage{booktabs} 

\usepackage{bussproofs}
\usepackage{multicol}
\usepackage[mathletters]{ucs}
\usepackage[utf8x]{inputenc}
\usepackage{here}
\PassOptionsToPackage{hyphens}{url}\usepackage{hyperref}
\usepackage{syntax}
\usepackage{amssymb}
\usepackage{amsmath}
\usepackage{xcolor}
\usepackage{tcolorbox}

\EnableBpAbbreviations

\newcommand*{\ppm}[3]{#1 \approx^{#2} #3 \Downarrow}
\newcommand*{\pdm}[2]{#1 \approx #2 \Downarrow}
\newcommand*{\mfun}[4]{#1 \sim^{#2}_{#3} #4 \Downarrow}
\newcommand*{\matom}[3]{#1 \sim_{#2} #3}
\newcommand*{\ev}[2]{#1, #2 \Downarrow}
\newcommand*{\set}[1]{\{#1\}}
\newcommand*{\cons}{:}
\newcommand*{\none}{\mathord{\texttt{none}}}
\newcommand*{\some}{\mathop{\texttt{some}}}
\newcommand*{\opt}{\mathop{\texttt{opt}}}

\newcommand*{\backgray}[1]{\colorbox[gray]{0.9}{\color[gray]{0.0}{#1}}}

\let\emptyset\varnothing

\begin{document}

\setlength{\pdfpageheight}{\paperheight}
\setlength{\pdfpagewidth}{\paperwidth}

\title{Loop Patterns: Extension of Kleene Star Operator for More Powerful Pattern Matching against Arbitrary Data Structures}

\author{Satoshi Egi}
\affiliation{%
  \institution{Rakuten Institute of Technology}
  \country{Japan}
}

\begin{abstract}
The Kleene star operator is an important pattern construct for representing a pattern that repeats multiple times.
Due to its simplicity and usefulness, it is imported into various pattern-matching systems other than regular expressions.
For example, Mathematica has a similar pattern construct called the repeated pattern.
However, they have the following limitations:
(i) We cannot change the pattern repeated depending on the current repeat count, and
(ii) we cannot apply them to arbitrary data structures such as trees and graphs other than lists.
This paper proposes the \emph{loop patterns} that overcome these limitations.
This paper presents numerous working examples and formal semantics of the loop patterns.
The examples in this paper are coded in the Egison programming language, which features the customizable non-linear pattern-matching facility for non-free data types.
\end{abstract}




\maketitle

\section{Introduction}\label{intro}


The Kleene star operator~\cite{hopcroft2013introduction}, sometimes called the repeated pattern~\cite{wolframRepeated}, is an important pattern construct for representing a pattern that repeats multiple times.
It allows us to represent patterns such as \texttt{[a,a,...]} (repetition of \texttt{a}), \texttt{[a,b,a,b,...]} (repetition of \texttt{[a,b]}), and \texttt{[a,b,a,a,b,...]} (repetition of \texttt{a} or \texttt{b}) for lists.
Due to its simplicity and usefulness, it is implemented in various pattern-matching systems.
For example, Mathematica~\cite{wolframRepeated} and Racket~\cite{tobin2011extensible} have the repeated pattern.
Domain specific languages such as Parsing expression grammars~\cite{ford2004parsing} and graph query languages such as Cypher~\cite{neo4jManual} and Gremlin~\cite{rodriguez2015gremlin} also have the Kleene star like operator.


However, the repeated patterns have the following two limitations.

\begin{enumerate}
\item We cannot change the content of the pattern repeated depending on the repeat count.
\item We can apply the repeated pattern only for lists.
\end{enumerate}


The first limitation becomes a problem, for example, when we consider the pattern that matches with lists of lists that the $n$-th list contains $n$ elements such as \texttt{[[a],[a,a]]} and \texttt{[[a],[a,a],[a,a,a]]}.
The repeated patterns cannot represent such a pattern because they do not provide users a method for referring to the current repeat count in the pattern.
The repeated patterns also does not allow us to refer to the values bound to the patterns repeated by $i-1$ times when we pattern-match the $n$-th repeat of the pattern repeated.
Consequently, we cannot describe a pattern that matches lists whose elements are in sequence such as \texttt{[1,2,3]} and \texttt{[2,3,4]} with the repeated patterns.


The second limitation is crucial for the general-purpose programming languages.
The reason for the second limitation is that a pattern repeated is always appended to the tail of the pattern.
For that reason, the repeated patterns can be applied only to collection data such as lists and multisets.~\footnote{Strings, which are the target of regular expressions, are a list of characters.}
As a result, for example, we cannot use the repeated patterns to traverse the ancestors or descendants of a tree node.




This paper presents the loop patterns that overcome the above two limitations.
We show the numerous sample programs of the loop patterns coded in the Egison programming language, which features the user-customizable non-linear pattern matching with backtracking for arbitrary user-defined data types~\cite{egison,egisonAplas}.

\section{Related Work}

There are tremendous amount of research on pattern-matching extensions~\cite{viewsWeb}.
This section compares these studies with our proposal.

Some of these studies allow users to define pattern-matching methods for each pattern constructor~\cite{wadler1987views,erwig1996active,tullsen2000first,antoy2010functional}.
It enables users to customize a pattern-matching method for arbitrary data types by themselves.
However, these proposals do not discuss on the repeated patterns.
For example, active patterns~\cite{erwig1996active,syme2007extensible} applies its pattern-matching facility to graphs~\cite{erwig1997functional}.
However, they use recursive functions for traversing graphs.
These proposals that provide a customizable pattern-matching facility do not support non-linear pattern matching with
multiple results except for Egison~\cite{egisonAplas}.
The repeated patterns are powerful especially when it is combined with non-linear pattern matching with backtracking.
It may be the reason that these proposals does not discuss the extension of the repeated patterns.

There is another approach, regular-tree expressions that extends regular expressions for handling trees.
Recursively defined patterns proposed in the \textsl{trx} regular-tree expression language~\cite{bagrak2004trx} overcomes the second limitation mentioned in Section~\ref{intro}, and enables us to apply repeated patterns to other than lists.
However, they still suffer from the first limitation.
The reason is because the recursively defined patterns do not provide a method for managing the repeat count.
The advantage of the loop patterns over the recursively defined patterns is its ability to refer to the repeat count in patterns.

\section{Pattern Matching of Egison}\label{egison}

\begin{figure}
\begin{grammar}
<expr> ::= <bool>                                       \hfill (boolean)
  \alt <number>                                         \hfill (number)
  \alt <ident>[`\_'<expr>]*                             \hfill (variable)
  \alt `<' <Ident> <expr>* `>'                          \hfill (algebraic data)
  \alt `[' <expr>* `]'                                  \hfill (tuple)
  \alt `{' <expr>* `}'                                  \hfill (collection)
  \alt `{|' <key-value>* `|}'                           \hfill (hash)
  \alt `(lambda [' `\$'<ident>* `]' <expr> `)'          \hfill (function)
  \alt `(' <expr> <expr>* `)'                           \hfill (function application)
  \alt `(if' <expr> <expr> <expr> `)'                   \hfill (if)
  \alt `(letrec {' <binding>* `}' <expr> `)'            \hfill (letrec)
  \alt `(match-all' <expr> <expr> `[' <pattern> <expr> `])' \hfill (match-all)
  \alt `something'                                  \hfill (something build-in matcher)
  \alt `(matcher \{' <matcher-clause>* `\})'           \hfill (matcher)

<key-value> :: `[' <number> <expr> `]' \hfill (hash key and value)

<binding> ::= `[' `\$'<ident> <expr> `]' \hfill (binding)

<pattern> ::= `\_' \hfill (wildcard)
  \alt `\$'<ident>[`\_'<expr>]*     \hfill (pattern variable)
  \alt `,'<expr>                      \hfill (value pattern)
  \alt `<' <ident> <pattern>* `>'      \hfill (inductive pattern)
  \alt `(|' <pattern>* `)'      \hfill (or-pattern)
  \alt `(&' <pattern>* `)'      \hfill (and-pattern)
  \alt `!' <pattern>      \hfill (not-pattern)
  \alt `(loop' `\$'<ident> <index-range> <pattern> <pattern>`)' \hfill (loop pattern)

<index-range> ::= `[' <expr> <expr> <pattern> `]' \hfill (index range)
\end{grammar}
  \caption{Formal grammar of Egison}
  \label{fig:syntax}
\end{figure}

Before starting the discussion on the loop patterns, this section gives a brief overview of our language focusing on its pattern-matching facility.
Our language features the user-customizable efficient non-linear pattern matching with backtracking.
The loop pattern is even more powerful when combined with this feature.

Figure~\ref{fig:syntax} shows the formal grammar of our language.
Our language is a Scheme-based language with lazy evaluation strategy.
The \texttt{match-all} and \texttt{match} expressions are syntax for pattern matching.
This section explains the usage of these expressions.
In Figure~\ref{fig:syntax}, \emph{ident} stands for an identifier that begins with an lowercase letter, whereas \emph{Ident} stands for an identifier that begins with an uppercase letter.

Our language uses four kinds of parenthesis in addition to ``\texttt{(}'' and ``\texttt{)}'', which denote function applications.
``\texttt{<}'' and ``\texttt{>}'' are used to apply pattern and data constructors.
In our language, the name of a data constructor starts with uppercase, whereas the name of a pattern constructor starts with lowercase.
``\texttt{[}'' and ``\texttt{]}'' are used to build a tuple.
``\verb|{|'' and ``\verb|}|'' are used to denote a collection.
``\verb!{|!'' and ``\verb!|}!'' are used to denote a hash table.


The following program is a sample of the \texttt{match-all} expression for handling pattern matching with multiple results.
In this paper, we show the evaluation result of a program in the comment that follows the program.
``\texttt{;}'' is the inline comment delimiter of our language.

{\footnotesize
\begin{verbatim}
(match-all {1 2 3} (list integer) [<join $xs $ys> [xs ys]])
; {[{} {1 2 3}] [{1} {2 3}] [{1 2} {3}] [{1 2 3} {}]}
\end{verbatim}
}

\texttt{match-all} is composed of an expression called \textit{target}, \textit{matcher}, and \textit{match clause}, which consists of a \textit{pattern} and \textit{body expression}.
The \texttt{match-all} expression evaluates the body of the match clause for each pattern-matching result and returns a collection that contains all results.
In the above code, we pattern-match the target \verb|{1 2 3}| as a list of integers using the pattern \texttt{<join \$xs \$ys>}.
\texttt{(list integer)} is a matcher to pattern-match the pattern and target as a list of integer.
The pattern is constructed using the \texttt{join} pattern constructor.
\texttt{\$xs} and \texttt{\$ys} are called \textit{pattern variables}.
We can use the result of pattern matching referring to them.
In the sample program, given a \texttt{join} pattern, \texttt{(list integer)} tries to divide the target collection into two collections.
The collection \verb|{1 2 3}| is thus divided into two collections by four ways.

Our language can handle non-linear patterns.
In our language, a pattern is examined from left to right in order, and the binding to a pattern variable can be referred to in its right side of the pattern.
In the following example, the pattern variable \texttt{\$p} is bound to one of prime numbers.
After that, the pattern ``\texttt{,(+ p 2)}'' is examined.
A pattern that begins with ``\texttt{,}'' is called a \textit{value pattern}.
The expression following ``\texttt{,}'' can be any kind of expressions.
The value patterns match with the target data if the target is equal to the content of the pattern.
Therefore, after successful pattern matching, \texttt{\$p} is bound to the first element of a prime twin.
Also note that, \texttt{match-all} can handle pattern matching that may yield infinitely many results thanks to the lazy evaluation strategy.

{\footnotesize
\begin{verbatim}
(define $twin-primes
  (match-all primes (list integer)
    [<join _ <cons $p <cons ,(+ p 2) _>>> [p (+ p 2)]]))

(take 6 twin-primes) ; {[3 5] [5 7] [11 13] [17 19] [29 31] [41 43]}
\end{verbatim}
}

Egison supports non-linear patterns instead of pattern guards because of its efficiency.
Implementation of efficient backtracking for non-linear patterns is easier than that for pattern guards that enumerate all matches before filtering the results using conditions described using pattern guards.
For example, the time complexity of the following pattern matching is $O(n^2)$.
This pattern matching checks weather the target collection contains an identical triple or not.
\texttt{between} is a function that return a collection consisting of the integers between the first and second arguments.
Therefore, the target collection contains no identical triple in this case.

{\footnotesize
\begin{verbatim}
(match-all (between 1 n) (multiset integer)
  [<cons $x <cons ,x <cons ,x _>>> x])
; returns {} in O(n^2)
\end{verbatim}
}

\noindent On the other hand, the time complexity of the following Curry program is $O(n^3)$.

{\footnotesize
\begin{verbatim}
insert x [] = [x]
insert x (y:ys) = x:y:ys ? y:(insert x ys)

tri (insert x (insert x (insert x _))) = "Matched"
tri _ = "Not matched"

tri [1,2,3,4,5,6,7,8,9,10]  -- returns "Not matched" in O(n^3) time
\end{verbatim}
}

\noindent The reason is because Curry transforms non-linear patterns into pattern guards as follows.
It enumerates all $\binom{n}{3}$ candidates before filtering them.

{\footnotesize
\begin{verbatim}
tri' (insert x (insert y (insert z _))) | x == y && x == z = "Matched"
tri' _ = "Not matched"
\end{verbatim}
}

There is another primitive syntax called \texttt{match} expression.
The major difference from \texttt{match-all} is that it can take multiple match clauses.
It tries pattern matching starting from the head of the match clauses, and tries the next clause if it fails.
Therefore, it is useful when we want to write conditional branching.
Another difference is that \texttt{match} only evaluates the first result of the results of the pattern matching, while \texttt{match-all} returns a collection of all results.

The \texttt{matcher} expression is a syntax to define a matcher.
Users can define the pattern-matching method for each data type using the \texttt{matcher} expression.
The usage of the \texttt{matcher} expression is explained in~\cite{egisonAplas}.
Extensibility of pattern matching is the core feature of our language.

All the matchers that appears in this paper including \texttt{integer}, \texttt{string}, \texttt{list}, \texttt{multiset}, and \texttt{set} are defined in our language using the \texttt{matcher} expression.
However, in reading this paper, we can consider them as built-in matchers.

\section{Loop Patterns}\label{result}

This section presents the loop patterns that overcome the limitations of the repeated patterns.

\subsection{Usage of Loop Patterns}\label{usage}

This section explains the grammar and usage of the loop patterns.

First, let us consider pattern matching for enumerating all combinations of two elements from a target collection.
It can be written using \texttt{match-all} as follows.

{\footnotesize
\begin{verbatim}
(define $comb2
  (lambda [$xs]
    (match-all xs (list integer)
      [<join _ <cons $x_1 <join _ <cons $x_2 _>>>>
       {x_1 x_2}])))

(comb2 {1 2 3 4}) ; {{1 2} {1 3} {2 3} {1 4} {2 4} {3 4}}
\end{verbatim}
}

Our language allows users to append indices to a pattern variable as \verb|$x_1| and \verb|$x_2| in the above sample.
They are called \textit{indexed variables} and represent $x_1$ and $x_2$ in mathematical expressions.
The expression after `_' must be evaluated to an integer and is called an \textit{index}.
We can append as many indices as we want.
When a value is bound to an indexed pattern variable \verb|$x_i|, the system initiates a hash if \verb|x| is not bound to a hash, and bind it to \texttt{x}.
If \verb|x| is already bound to a hash, the new key and value is added to the hash.

Next, let us consider pattern matching for enumerating all combinations of \emph{three} elements from the target collection.
It can be written by only modifying the above program a bit as follows.

{\footnotesize
\begin{verbatim}
(define $comb3
  (lambda [$xs]
    (match-all xs (list integer)
      [<join _ <cons $x_1 <join _ <cons $x_2 <join _ <cons $x_3 _>>>>>>
       {x_1 x_2 x_3}])))

(comb3 {1 2 3 4}) ; {{1 2 3} {1 2 4} {1 3 4} {2 3 4}}
\end{verbatim}
}

Now, let us generalize \texttt{comb2} and \texttt{comb3}.
The loop patterns can be used for that purpose.\footnote{This generalization can be done by also the repeated patterns. We show pattern matching that can be done only by the loop patterns in the next section.}

{\footnotesize
\begin{verbatim}
(define $comb
  (lambda [$n $xs]
    (match-all xs (list integer)
      [(loop $i [1 {n} _]
         <join _ <cons $x_i ...>>
         _)
       (map (lambda [$i] x_i) (between 1 n))])))

(comb 2 {1 2 3 4}) ; {{1 2} {1 3} {2 3} {1 4} {2 4} {3 4}}
(comb 3 {1 2 3 4}) ; {{1 2 3} {1 2 4} {1 3 4} {2 3 4}}
\end{verbatim}
}

As shown in Figure~\ref{fig:syntax}, the loop pattern takes an \emph{index variable}, \emph{index range}, \emph{repeat pattern}, and \emph{end pattern} as arguments.
An index variable is a variable to hold the current repeat count.
An index range specifies the range where the index variable moves.
A repeat pattern is a pattern repeated when the index variable is in the index range.
An end pattern is a pattern expanded when the index variable get out of the index range.

Inside loop patterns, we can use the \emph{ellipsis pattern} (\texttt{...}).
The repeat pattern or the end pattern is expanded at the location of the ellipsis pattern.
When the repeat pattern is expanded replacing the ellipsis pattern, the index variable is incremented.
When the end pattern is expanded replacing the ellipsis pattern, the end number pattern is pattern-matched with the value bound to the index variable at that time.

The index range is composed of a \emph{start number}, \emph{end numbers}, and \emph{end number pattern}.
A start number specifies the initial value of the index variable.
End numbers are a sorted list of integers.
When the current value of index variable is not in the end numbers, the ellipsis pattern is replaced only with the repeat pattern.
When the index variable reaches one of the end numbers, the ellipsis pattern is replaced with both of the end pattern and the repeat pattern.
When the index variable reaches the last end number, the ellipsis pattern is replaced only with the end pattern.

In the above sample, the start number is \texttt{1} and the end numbers are \verb|{n}|.
In this case, the end number can be only \texttt{n}.
Therefore, the index variable moves from \texttt{1} to \texttt{n}.

If we use a loop pattern we can redefine \texttt{comb2} as follows.
In this sample, the index range is \verb|[1 {2} _]|.
The start number is \verb|1|, which means the index variable \verb|i| starts from \verb|1|.
The end numbers are \verb|{2}|, which means the index variable \verb|i| moves to \verb|2| and the pattern is repeated twice.
The end number pattern is the wildcard \verb|_|.
In this case, the end number can be only \texttt{2}, therefore, we do not need to bind it to a pattern variable.

{\footnotesize
\begin{verbatim}
(define $comb2
  (lambda [$xs]
    (match-all xs (list integer)
      [(loop $i [1 {2} _]
         <join _ <cons $x_i ...>>
         _)
       {x_1 x_2}])))

(comb2 {1 2 3 4})
; {{1 2} {1 3} {2 3} {1 4} {2 4} {3 4}}
\end{verbatim}
}

The following program is an example that utilize both of end numbers and an end number pattern.
The \texttt{comb2or3} function returns all combinations of two or three elements from the argument collection.
\texttt{2} or \texttt{3} is bound to \verb|$n| because the end number pattern \verb|$n| and the index variable that reached the end numbers \texttt{2} or \texttt{3} are pattern-matched.


{\footnotesize
\begin{verbatim}
(define $comb2or3
  (lambda [$xs]
    (match-all xs (list integer)
      [(loop $i [1 {2 3} $n]
         <join _ <cons $x_i ...>>
         _)
       (map (lambda [$i] x_i) (between 1 n))])))

(comb2or3 {1 2 3 4})
; {{1 2} {1 3} {2 3} {1 4} {2 4} {3 4} {1 2 3} {1 2 4} {1 3 4} {2 3 4}}
\end{verbatim}
}

We prepared several syntax sugar for the index range as follows.
It allows us to omit end numbers, an end number pattern, or both.

{\footnotesize
\begin{verbatim}
[ <start-num> ].              ;=> [ <start-num> (from <start-num>) _ ]
[ <start-num> <end-numbers> ] ;=> [ <start-num> <end-numbers> _ ]
[ <start-num> <end-number> ]  ;=> [ <start-num> {<end-number>} _ ]
[ <start-num> <end-pattern> ] ;=> [ <start-num> (from start-num>) <end-pattern> ]
\end{verbatim}
}

A start number cannot be omitted.
When end numbers are omitted, \texttt{(from <start number>)} is complemented.
\texttt{from} is a function that returns the infinite sequence of integers that starts from its argument.
If end numbers are not a collection but a single number, it is converted to the collection that contains only that end number.
When an end number pattern is omitted, the wildcard \verb|_| is complemented.

If we use these syntax sugar, we can rewrite the index range of \texttt{comb2} to \verb|[1 2]| as follows.

{\footnotesize
\begin{verbatim}
(define $comb2
  (lambda [$xs]
    (match-all xs (list integer)
      [(loop $i [1 2]
         <join _ <cons $x_i ...>>
         _)
       {x_1 x_2}])))

(comb2 {1 2 3 4})
; {{1 2} {1 3} {2 3} {1 4} {2 4} {3 4}}
\end{verbatim}
}

\subsection{Advantages of Loop Patterns}

The first characteristic of the loop patterns is that we explicitly name an index variable (\verb|$i| and \verb|$j| in the following example).
We can parameterize repeat patterns by a current repeat count by referring to the index variable.

The following program describes a pattern that matches with lists of lists whose $i$-th list contains $i$ elements.
Note that we can nest the loop patterns.
We refers to the index variable \verb|i| of the first loop pattern in the end number of the second loop pattern.

{\footnotesize
\begin{verbatim}
(match {{1} {2 2} {3 3 3} {4 4 4 4}} (list (list integer))
  {[(loop $i [1 $n]
     <cons (loop $j [2 i]
             <cons _ ...>
             <nil>)
           ...>
      <nil>)
    #t]
    [_ #f]})
; #t
\end{verbatim}
}

The pattern in the following program matches if the elements of a target list are in sequence.
We are referring to the value bound in the pattern previously repeated as \texttt{,(+ 1 x_(- i 1))}.
Note that index variables play the crucial roles in this pattern. 
In Mathematica~\cite{wolframRepeated} and Racket~\cite{tobin2011extensible}, the values bound to a repeated pattern are stored as elements of a collection.
However, they are stored in a hash table in our language.
It allows us to easily access the value bound to the pattern variables in the loop pattern.
In this example, the and-pattern is used as an as-pattern.

{\footnotesize
\begin{verbatim}
(match {1 2 3 4 5} (list integer)
  {[<cons $x_1
     (loop $i [1 $n]
       <cons (& ,(+ 1 x_(- i 1)) $x_i)
        ...>
     <nil>)>
    #t]
    [_ #f]})
; #t
\end{verbatim}
}

In Section~\ref{demo-n-queens} and~\ref{demo-graph}, we show more interesting samples utilizing this characteristic of the loop patterns.

The second characteristic of the loop patterns is that we can place a ellipsis pattern in a place where we like.
This feature allows us to use the loop patterns for arbitrary user-defined data structures, not only for collections.
For example, we can apply the loop patterns for trees (Section~\ref{demo-tree}).

\section{Examples}\label{demo}

This section demonstrates working examples of the loop patterns.

\subsection{N-queens Solver}\label{demo-n-queens}

\begin{figure}
  \begin{center}
    
{\footnotesize
\begin{verbatim}
(define $four-queens
  (match-all {1 2 3 4} (multiset integer)
    [<cons $a_1
      <cons (& !,(- a_1 1) !,(+ a_1 1) $a_2)
       <cons (& !,(- a_1 2) !,(+ a_1 2) !,(- a_2 1) !,(+ a_2 1) $a_3)
        <cons (& !,(- a_1 3) !,(+ a_1 3) !,(- a_2 2) !,(+ a_2 2) !,(- a_3 1) !,(+ a_3 1) $a_4)
         <nil>>>>>
     {a_1 a_2 a_3 a_4}]))

four-queens ; {{2 4 1 3} {3 1 4 2}}
\end{verbatim}
}
  \end{center}
  \caption{Four queens solver}
  \label{fig:four-queens}
\end{figure}

Figure~\ref{fig:four-queens} shows the four queens solver.
$n$-queen problem is the problem of the placing $n$ chess queens on an $n \times n$ board such that no queen is able to attack any of the others.
The chess queen can attack other chess pieces on the same row, column, and diagonal.

In Figure~\ref{fig:four-queens}, we represent the positions of the four queens with a list.
The number of the $n$-th element represents the row number of the queen of the $n$-th line.
The solution must be a rearrangement of the list \verb|{1 2 3 4}| because no two queens can be in the same line or row.
Therefore, we pattern-match a collection \verb|{1 2 3 4}| as a multiset of integers.
The requirement that all two queens must not share the same diagonal is represented with conditions $a_1 \pm 1 \neq a_2$, $a_1 \pm 2 \neq a_3$, $a_2 \pm 1 \neq a_3$, $a_1 \pm 3 \neq a_4$, $a_2 \pm 2 \neq a_4$, and $a_3 \pm 1 \neq a_4$.
\emph{Not-patterns} are used to represent $\neq$.
A not-pattern matches with a object if the object does not matches the pattern following after `\texttt{!}'.

Next, let us consider defining five-queens, six-queens, seven-queens, eight-queens solvers, and so on.
Patterns for these solvers have a similar form and it would be better to define the general $n$-queens solver at once than to define each of them.
The loop patterns can be used for that.

\begin{figure}
  \begin{center}
    
{\footnotesize
\begin{verbatim}
(define $n-queens
  (lambda [$n]
    (match-all (between 1 n) (multiset integer)
      [<cons $a_1
        (loop $i [2 n]
              <cons (loop $j [1 (- i 1)]
                          (& !,(- a_j (- i j)) !,(+ a_j (- i j)) ...)
                          $a_i)
                    ...>
              <nil>)>
      a])))

(n-queens 4) ; {{|[1 2] [2 4] [3 1] [4 3]|} {|[1 3] [2 1] [3 4] [4 2]|}}
\end{verbatim}
}
  \end{center}
  \caption{$n$-queens solver}
  \label{fig:n-queens}
\end{figure}

We can generalize \texttt{four-queens} to $n$-queens solver using loop patterns as Figure~\ref{fig:n-queens}.
A double loop pattern is used to express the pattern for $n$-queens solver.

If we use loop patterns in a pattern, the numbers of the pattern variables in the pattern can change by the parameter.
It is the reason why we introduced indexed variables.

This pattern is an example that can be described only by the loop patterns because
\begin{enumerate}
\item it refers to the index variable \verb|i| (the current repeat count) of the first loop pattern in the end number of the second loop pattern, and
\item it refers to the value bound in the previously repeated pattern as ``\texttt{,a_j}''.
\end{enumerate}

\subsection{Pattern Matching for Trees}\label{demo-tree}

\begin{figure}
  \begin{center}
    
{\footnotesize
\begin{verbatim}
(define $tree (algebraic-data-matcher {<leaf string> <node string (multiset tree)>}))

(define $tree-data
  <Node "Programming language"
    {<Node "Pattern-matching-oriented"
       {<Leaf "Egison">}>
     <Node "Functional language"
       {<Node "Strictly typed"
          {<Leaf "OCaml"> <Leaf "Haskell"> <Leaf "Curry"> <Leaf "Coq">}>
        <Node "Dynamically typed"
          {<Leaf "Egison"> <Leaf "Lisp"> <Leaf "Scheme"> <Leaf "Racket"> <Leaf "Clojure">}>
        }>
     <Node "Logic programming"
       {<Leaf "Prolog"> <Leaf "LiLFeS"> <Leaf "Curry">}>
     <Node "Object oriented"
       {<Leaf "C++"> <Leaf "Java"> <Leaf "Ruby"> <Leaf "Python"> <Leaf "OCaml">}>
     }>)
     
;; All categories that Egison belongs.
(match-all tree-data tree
  [(loop $i [1 $n]
         <node $c_i <cons ... _>>
         <leaf ,"Egison">)
   c]))
;{{|[1 "Programming language"] [2 "Pattern-matching-oriented"]|}
  {|[1 "Programming language"] [2 "Functional language"] [3 "Dynamically typed"]|}}
\end{verbatim}
}
  \end{center}
  \vspace{-2mm}
  \caption{Pattern matching for trees}
  \label{fig:tree}
  \vspace{-4mm}
\end{figure}

Figure~\ref{fig:tree} demonstrates pattern matching for trees.
A matcher for trees is defined using the \texttt{algebraic-data-matcher} expression~\cite{egi2014non}.
Trees have two data constructors, \texttt{leaf} and \texttt{node}.
\texttt{leaf} obtains one argument and it is pattern-matched as a string.
\texttt{node} obtains two arguments, and the first and second argument are pattern-matched as a string and  a multiset of trees, respectively.

In Figure~\ref{fig:tree}, \texttt{tree-data} defines a category tree of programming languages.
For example, \texttt{"Egison"} belongs to the \texttt{"Pattern-matching-oriented"} category, and the \texttt{"Dynamically typed"} sub-category of the \texttt{"Functional programming"} category.

The \texttt{match-all} expression in Figure~\ref{fig:tree} enumerates all categories to which \texttt{"Egison"} belongs.
A loop pattern is necessary to describe a pattern for this purpose because leaves can be appear in arbitrary depth.

This pattern matching is an illustrative example to show the importance of the ellipsis pattern.
The characteristic that we can choose the position where the repeat pattern is expanded allows us to apply the loop patterns to trees.

XML path language~\cite{berglund2003xml} is a domain specific language for pattern matching against XML trees.
In XML path language, we use the built-in \texttt{ancestor} command to enumerates all ancestors of a node.
However, in our language, we can do the same thing by just combining the loop patterns and a few simple pattern constructors.

\subsection{Pattern Matching for Graphs}\label{demo-graph}

\begin{figure}
  \begin{center}

{\footnotesize
\begin{verbatim}
(define $graph (set edge))
(define $edge (algebraic-data-matcher {<edge integer integer>}))
(define $graph-data {<Edge 1 4> <Edge 2 1> <Edge 3 1> <Edge 3 2> <Edge 4 3> <Edge 5 1> <Edge 5 4>})

;; The shortest path paths from 's' to 'e'.
(car (let {[$s 1] [$e 2]}
       (match-all graph-data graph
         [(let {[$x_1 s]}
           (loop $i [2 $n]
             <cons <edge ,x_(- i 1) $x_i> ...>
             <cons <edge ,x_(- n 1) (& ,e $x_n)> _>))
          (map (lambda [$i] x_i) (between 1 n))])))
; {1 4 3 2}
\end{verbatim}
  }

  \end{center}
  \caption{Pattern matching for graphs as a set of edges}
  \vspace{-2mm}
  \label{fig:graph1}
  \vspace{-3mm}
\end{figure}

In Figure~\ref{fig:graph1}, a graph is pattern-matched as a set of edges.
This pattern is an example that can be described only by the loop pattern.
The reason is because it refers to the value bound in the previously repeated pattern as ``\verb|,x_(- i 1)|''.

We can use the \texttt{let} expression inside a pattern.
A \texttt{let} expression inside a pattern is called a \emph{let pattern}.
In this pattern, the let pattern is effectively used to bind \texttt{s} to \verb|$x_1|.
Thanks to this let pattern, we do not need the special treatment for \verb|$x_1|.

\begin{figure}
  \begin{center}

{\footnotesize
\begin{verbatim}
(define $station string)
(define $price   integer)
(define $graph   (multiset [station (multiset [station price])]))

(define $graph-data
  {
   ["Tokyo"     {              ["Shinjuku" 200] ["Shibuya" 200] ["Mitaka" 390] ["Kinshicho" 160] ["Kitasenju" 220]}]
   ["Shinjuku"  {["Tokyo" 200]                  ["Shibuya" 160] ["Mitaka" 220] ["Kinshicho" 220] ["Kitasenju" 310]}]
   ["Shibuya"   {["Tokyo" 200] ["Shinjuku" 160]                 ["Mitaka" 310] ["Kinshicho" 220] ["Kitasenju" 310]}]
   ["Mitaka"    {["Tokyo" 390] ["Shinjuku" 220] ["Shibuya" 310]                ["Kinshicho" 470] ["Kitasenju" 550]}]
   ["Kinshicho" {["Tokyo" 160] ["Shinjuku" 220] ["Shibuya" 220] ["Mitaka" 470]                   ["Kitasenju" 220]}]
   ["Kitasenju" {["Tokyo" 220] ["Shinjuku" 310] ["Shibuya" 310] ["Mitaka" 550] ["Kinshicho" 220]                  }]
   })

;; List all routes that visit all cities exactly once and return to Tokyo.
(define $trips
  (match-all graph-data graph
    [<cons [,"Tokyo" <cons [$s_1 $p_1] _>]
           (loop $i [2 5]
             <cons [,s_(- i 1) <cons [$s_i $p_i] _>]
                   ...>
             <cons [,s_5 <cons [(& ,"Tokyo" $s_6) $p_6] _>]
                   _>)>
     [(sum (map (lambda [$i] p_i) (between 1 6)))
      s]]))
\end{verbatim}
  }

  \end{center}
  \vspace{-2mm}
  \caption{Pattern matching for graphs as an adjacency list}
  \label{fig:graph2}
  \vspace{-3mm}
\end{figure}

In Figure~\ref{fig:graph2}, a graph is pattern-matched as an adjacency list.
\texttt{graph-data} in Figure~\ref{fig:graph2} represents the railway network in Tokyo.
\texttt{"Tokyo"}, \texttt{"Shibuya"}, \texttt{"Mitaka"}, \texttt{"Kinshicho"}, and \texttt{"Kitasenju"} are names of the train stations in Tokyo.
The integers in \texttt{graph-data} are train fees.

The pattern in Figure~\ref{fig:graph2} lists all routes that visit all cities exactly once and return to Tokyo.
This pattern can be used for solving the traveling salesman problem.
It is also an example that can be described only by the loop patterns because it refers to the value bound in the previously repeated pattern as ``\verb|,s_(- i 1)|''.

There are several graph database query languages~\cite{perez2006semantics,rodriguez2015gremlin,neo4jManual}.
The advantages of our language against these query languages are as follows.

Our language does not focus on pattern matching for graphs.
However, we can represent various patterns by combining the loop patterns and a small number of simple pattern constructors.

\subsection{\texttt{take} and \texttt{drop}}

Pattern matching with backtracking allows us to define the basic functions for list programming more concisely than the traditional functional definitions.
The reason is because we can hide recursion to traverse the elements of lists by pattern matching.
For example, the \texttt{map} function can be defined as follows~\cite{egi2015egison}.
\texttt{something} in the following code is the only built-in matcher of our pattern-matching system.
\texttt{something} can handle only wildcards and pattern variables, and is used to bind a value to a pattern variable.

{\footnotesize
\begin{verbatim}
(define $map
 (lambda [$f $xs]
  (match-all xs (list something)
   [<join _ <cons $x _>> (f x)])))
\end{verbatim}
}

The loop patterns are useful to simplify some basic functions with pattern matching.
For example, the \texttt{take} and \texttt{drop} functions are implemented using loop patterns as follows.

{\footnotesize
\begin{verbatim}
(define $take
 (lambda [$n $xs]
  (match xs (list something)
   {[(loop $i [1 n] <cons $x_i ...> _) (map (lambda [$i] x_i) (between 1 n))]
    [_ xs]})))

(define $drop
 (lambda [$n $xs]
  (match xs (list something)
   {[(loop $i [1 n] <cons _ ...> $ys) ys]
    [_ {}]})))
\end{verbatim}
}

\section{Discussion}\label{discussion}

This section discusses the reasonableness of the design of the loop patterns.

\subsection{Necessity of End Numbers}

If we write \verb|[1 ,2]| instead of \verb|[1 2]|, and \verb![1 (& (| ,2 ,3) $n)]! instead of \verb|[1 {2 3} $n]|, we can remove the end numbers from the index range.
It makes the syntax of the loop patterns simpler.
However, cases in which the loop pattern with this syntax cannot be executed correctly exist.

For example, the program for the $n$-queens solver causes a runtime error if we replace the index range from \texttt{[1 (- i 1)]} to \texttt{[1 ,(- i 1)]}.
The reason is because the index variable \texttt{j} does not stop at \texttt{i}$-$\texttt{1} and moves to \texttt{i}.
Therefore, the program causes an unbound variable error when the system refers to the value of \verb|a_j| inside the value pattern.

If the index range is \verb|[1 2]|,  the index variable moves \verb|1| and \verb|2|.
However, if the index range is \verb|[1 ,2]|,  the index variable cannot stop at \verb|2| and moves \verb|1|, \verb|2|, and \verb|3|.
This is because the system does not detect the integers greater than 2 do not match with the pattern ``\verb|,2|''.
We allow any patterns for the end pattern and there are cases that value patterns are inside an or-pattern, and-pattern, or more complicated pattern.
However, a special treatment for value patterns makes the implementation of the loop pattern complicated and unnatural.
This is the reason why the end pattern is necessary.

\subsection{Computational Complexity of Loop Patterns}

A loop pattern is expanded only when it is necessary.
Therefore, the time complexity of pattern matching using loop patterns is completely same with pattern matching written not using them.

The computational complexity of Egison pattern matching is discussed in~\cite{egisonAplas} in detail.

\section{Formal Semantics}\label{formal-semantics}

\begin{figure}[!t]
  \scriptsize
  \begin{gather*}
  \intertext{Evaluation of \texttt{matcher} and \texttt{match-all}:}
    \AXC{}
    \UIC{$\ev{\Gamma}{\texttt{(matcher $\texttt{[$pp_i$ $M_i$ $\texttt{[$dp_j$ $N_j$]}_j$]}_i$)}} ([pp_i, M_i, [dp_j, N_j]_j]_i,\Gamma)$}
    \DP
    \\
    \AXC{$\ev{\Gamma}{M} v$}
    \AXC{$\ev{\Gamma}{N} m$}
    \AXC{$[[[\matom{p}{m}{v}],\Gamma,\emptyset\backgray{$,\epsilon$}]] \Rrightarrow [\Delta_i]_i $}
    \AXC{$\ev{\Gamma \cup \Delta_i}{L} v_i \quad (\forall i)$}
    \QIC{$\ev{\Gamma}{\texttt{(match-all $M$ $N$ [$p$ $L$])}} [v_i]_i$}
    \DP
  \intertext{Matching states:}
    \AXC{}
    \UIC{$\epsilon \rightarrow \none, \none, \none$}
    \DP
    \quad
    \AXC{}
    \UIC{$(\epsilon, \Gamma, \Delta\backgray{$, \Lambda$}):\vec{s} \rightarrow (\some \Delta), \none, (\some \vec{s})$}
    \DP
    \\
    \colorbox[gray]{0.9}{\color[gray]{0.0}{
    \AXC{$\ev{\Gamma}{M} n$}
    \AXC{$\ev{\Gamma}{N} {\{e_i\}_i}$}
    \AXC{$((\matom{\texttt{...}}{m}{v}) \cons \vec{a}, \Gamma, \Delta, (\set{\texttt{i} \mapsto n-1},\{e_i\}_i, (p_1, p_2, p_3)) \cons \Lambda):\vec{s} \rightarrow \none, (\some \vec{s'}), (\some \vec{s})$}
    \TIC{$((\matom{\texttt{(loop \$i [$M \ N \  p_1$] $p_2 \  p_3$)}}{m}{v}) \cons \vec{a}, \Gamma, \Delta, \Lambda):\vec{s} \rightarrow \none, (\some \vec{s'}), (\some \vec{s})$}
    \DP
    }}
    \\
    \colorbox[gray]{0.9}{\color[gray]{0.0}{
    \AXC{$i \neq e_1$}
    \UIC{$((\matom{\texttt{...}}{m}{v}) \cons \vec{a}, \Gamma, \Delta, (\set{\texttt{i} \mapsto i},\{e_i\}_i, (p_i)_i) \cons \Lambda):\vec{s} \rightarrow \none, (\some [((\matom{p_2}{m}{v}) \cons \vec{a}), \Gamma, \Delta, (\set{\texttt{i} \mapsto i+1},\{e_i\}_{i}, [p_i]_i) \cons \Lambda]), (\some \vec{s})$}
    \DP
    }}
    \\
    \colorbox[gray]{0.9}{\color[gray]{0.0}{
    \AXC{$i = e_1$}
    \UIC{$((\matom{\texttt{...}}{m}{v}) \cons \vec{a}, \Gamma, \Delta, (\set{\texttt{i} \mapsto i},\{e_i\}_i, (p_i)_i) \cons \Lambda):\vec{s} \rightarrow \none, (\some \begin{bmatrix}((\matom{p_1}{\texttt{something}}{i}) \cons (\matom{p_3}{m}{v}) \cons \vec{a}, \Gamma, \Delta, \Lambda), \\ ((\matom{p_2}{m}{v}) \cons \vec{a}, \Gamma, \Delta, (\set{\texttt{i} \mapsto i+1},\{e_i\}_{i'}, (p_i)_i) \cons \Lambda) \end{bmatrix}), (\some \vec{s})$}
    \DP
    }}
    \\
    \colorbox[gray]{0.9}{\color[gray]{0.0}{
    \AXC{$i = e_1$}
    \UIC{$((\matom{\texttt{...}}{m}{v}) \cons \vec{a}, \Gamma, \Delta, (\set{\texttt{i} \mapsto i},\{e_1\}, (p_i)_i) \cons \Lambda):\vec{s} \rightarrow \none, (\some [((\matom{p_1}{\texttt{something}}{i}) \cons (\matom{p_3}{m}{v}) \cons \vec{a}), \Gamma, \Delta, \Lambda]), (\some \vec{s})$}
    \DP
    }}
    \\
    \AXC{$\mfun{p}{\Gamma \cup \Delta\backgray{$ \cup \sum_i \Delta'_i$}}{m}{v} [\vec{a}_i]_i, \Delta''$}
    \UIC{$((\matom{p}{m}{v}) \cons \vec{a}, \Gamma, \Delta\backgray{$, \{ \Delta'_i, \vec{e}_i, \vec{p}_i \}_{i}$}):\vec{s} \rightarrow \none, (\some [\vec{a}_i + \vec{a}, \Gamma, \Delta \cup \Delta''\backgray{$, \{ \Delta'_i, \vec{e}_i, \vec{p}_i \}_{i}$}), (\some \vec{s})$}
    \DP
    \\
    \AXC{$\vec{s}_i \rightarrow \opt \Gamma_i, \opt \vec{s'}_i, \opt \vec{s''}_i \quad (\forall i)$}
    \UIC{$[\vec{s}_i]_i \Rightarrow \sum_i (\opt \Gamma_i), \sum_i (\opt \vec{s'}_i) + \sum_i (\opt \vec{s''}_i)$}
    \DP
    \qquad
    \AXC{$\mathstrut$}
    \UIC{$\epsilon \Rrightarrow \epsilon$}
    \DP
    \quad
    \AXC{$\vec{\vec{s}} \Rightarrow \vec{\Gamma}, \vec{\vec{s'}}$}
    \AXC{$\vec{\vec{s'}} \Rrightarrow \vec{\Delta}$}
    \BIC{$\vec{\vec{s}} \Rrightarrow \vec{\Gamma} + \vec{\Delta}$}
    \DP
  \intertext{Matching atoms:}
    \AXC{$\mathstrut$}
    \UIC{$\mfun{\texttt{\$x}}{\Gamma}{\texttt{something}}{v} [\epsilon], \set{x \mapsto v}$}
    \DP
    \quad
    \AXC{$\ppm{pp}{\Gamma}{p} \textbf{fail}$}
    \AXC{$\mfun{p}{\Gamma}{(\vec{\phi},\Delta)}{v} \vec{\vec{a}}, \Gamma'$}
    \BIC{$\mfun{p}{\Gamma}{((pp,M,\vec{\sigma}) \cons \vec{\phi},\Delta)}{v} \vec{\vec{a}}, \Gamma'$}
    \DP
    \\
    \AXC{$\ppm{pp}{\Gamma}{p} [p'_i]_i, \Delta'$}
    \AXC{$\pdm{dp}{v} \textbf{fail}$}
    \AXC{$\mfun{p}{\Gamma}{((pp,M,\vec{\sigma}) \cons \vec{\phi},\Delta)}{v} \vec{\vec{a}}, \Gamma'$}
    \TIC{$\mfun{p}{\Gamma}{((pp,M,(dp,N) \cons \vec{\sigma}) \cons \vec{\phi},\Delta)}{v} \vec{\vec{a}}, \Gamma'$}
    \DP
    \\
    \AXC{$\ppm{pp}{\Gamma}{p} [p'_j]_j, \Delta'$}
    \AXC{$\pdm{dp}{v} \Delta''$}
    \AXC{$\ev{\Delta \cup \Delta' \cup \Delta''}{N} [[v'_{ij}]_j]_i$}
    \AXC{$\ev{\Delta}{M} [m'_j]_j$}
    \QIC{$\mfun{p}{\Gamma}{((pp,M,(dp,N) \cons \vec{\sigma}) \cons \vec{\phi},\Delta)}{v} [[\matom{p'_j}{m'_j}{v'_{ij}}]_j]_i, \emptyset$}
    \DP
  \intertext{Pattern matching on patterns:}
    \AXC{$\mathstrut$}
    \UIC{$\ppm{\texttt{\$}}{\Gamma}{p} [p], \emptyset$}
    \DP
    \quad
    \AXC{$\ev{\Gamma}{M} v$}
    \UIC{$\ppm{\texttt{,\$y}}{\Gamma}{\texttt{,$M$}} \epsilon, \set{y \mapsto v}$}
    \DP
    \quad
    \AXC{$\ppm{pp_i}{\Gamma}{p_i} \vec{p}_i, \Gamma_i \quad (\forall i)$}
    \UIC{$\ppm{\texttt{<C $pp_1 \ldots pp_n$>}}{\Gamma}{\texttt{<C $p_1 \ldots p_n$>}} \sum_i \vec{p}_i, \bigcup_i \Gamma_i$}
    \DP
  \intertext{Pattern matching on data:}
    \AXC{$\mathstrut$}
    \UIC{$\pdm{\texttt{\$z}}{v} \set{z \mapsto v}$}
    \DP
    \qquad
    \AXC{$\pdm{dp_i}{v_i} \Gamma_i \quad (\forall i)$}
    \UIC{$\pdm{\texttt{<C $dp_1 \ldots dp_n$>}}{\texttt{<C $v_1 \ldots v_n$>}} \bigcup_i \Gamma_i$}
    \DP
  \end{gather*}
  \caption{Formal semantics of the loop patterns}
  \label{fig:formal-semantics}
\end{figure}

Figure~\ref{fig:formal-semantics} shows the formal semantics of pattern matching of Egison including the loop patterns.
This formal semantics is based on the formal semantics presented in~\cite{egisonAplas}.
The rules newly added for handling the loop patterns are highlighted to make the difference from~\cite{egisonAplas} clear.

First, we explain the notations used in highlighted rules in Figure~\ref{fig:formal-semantics}.
$[a_i]_i$ denotes a list $[a_1,a_2,...]$.
$[[a_{ij}]_j]_i$ denotes a list of lists $[[a_{11},a_{12},...],[a_{21},a_{22},...]]$.
Similarly, $(a_i)_i$ denotes a tuple $(a_1,a_2,...)$.
A list of tuples is denoted as $[a_i, b_i]_i$ omitting the parenthesis.
$l_1 + l_2$ denotes a concatenation of lists $l_1$ and $l_2$.
$a \cons l$ has the same meaning with $[a] + l$.
$\epsilon$ denotes an empty list.
$\vec{x}$ for some metavariable $x$ is denotes a list of what $x$ denotes.

Matching states have a stack of loop contexts in addition to a stack of matching atoms, an environment, and an intermediate result of pattern matching.
$\Lambda$ denotes a stack of loop contexts.
A loop context consists of a binding information of an index variable, end numbers, and a tuple of an end number pattern, repeat pattern, and end pattern.
$p_1$, $p_2$, and $p_3$ denote an end number pattern, repeat pattern, and end pattern, respectively.
The stack of the loop contexts for an initial matching state is empty.

Four rules are newly added to the rules for matching states in order to handle the loop patterns and ellipsis patterns.
When the pattern of the top matching atom is a loop pattern, a new loop context is created and pushed to the stack of the loop contexts.
When the pattern of the top matching atom is an ellipsis pattern, the system refers to the top loop context to replace the ellipsis pattern to the end pattern or repeat pattern.

\section{Conclusion}\label{conclusion}

This paper proposed the loop patterns that overcome the limitations of the Kleene star operator and the repeated patterns and showed the several working examples that demonstrate the expressiveness of the loop patterns.
We showed that the loop pattern can be used to represent more expressive patterns for lists.
Furthermore, we also showed that the loop pattern can represent expressive patterns even for trees and graphs by only using a small number of simple pattern constructors.

We leave integration of our pattern-matching system with various query languages including SQL~\cite{chamberlin1974sequel}, XML path language~\cite{berglund2003xml} and graph query languages~\cite{perez2006semantics,rodriguez2015gremlin,neo4jManual} as future work.
These query languages are focusing on handling only their target data structures and have many built-in functions to handle various patterns.
On the other hand, our pattern-matching system allows users to describe various patterns for various data types in a unified way with a small number of pattern constructors and the loop patterns.
For this integration, we need to discuss an efficient execution method of our pattern-matching system and the loop patterns.

Research for finding simpler language constructs for constructing the loop patterns is also interesting because the semantics of the loop patterns presented in this paper is a bit complicated as a built-in language feature.

The loop pattern expands the range of algorithms we can describe concisely using the user-customizable efficient non-linear pattern-matching facility proposed in~\cite{egisonAplas}.
We hope this work leads to the further research of pattern matching as supportive evidence for the importance of pattern matching for more intuitive representation of algorithms.

\begin{acks}
I thank Kentaro Honda for giving me a program of the $n$-queen problem solver as an illustrative sample of the loop patterns.
I thank Yuichi Nishiwaki for reviewing the formal semantics of the loop patterns.
\end{acks}

\bibliographystyle{ACM-Reference-Format}
\bibliography{egison}

\end{document}